\documentclass[aps,prd,longbibliography,preprint,amsmath,amssymb,floatfix,dvipdfmx]{revtex4-1}

\usepackage{graphicx}

\usepackage{xcolor}
\usepackage{url}
\usepackage{bookmark}
\usepackage{lineno}

\usepackage{bm}
\usepackage{slashbox,multirow}

\newcommand{\be}{\begin{equation}}
\newcommand{\ee}{\end{equation}}
\newcommand{\eq}[1]{Eq.~(\ref{#1})}
\newcommand{\fig}[1]{Fig.~\ref{#1}}
\def\bea{\begin{eqnarray}}
\def\eea{\end{eqnarray}}

\def\bra{\langle}
\def\ket{\rangle}
\def\vq{{\bf q}}

\def\vk{{\bf k}}

\def\qp{{\bf q}_{\parallel}}

\begin{document}

\title{Theoretical perspectives on charge dynamics in high-temperature cuprate superconductors} 

\author{Hiroyuki Yamase}
\affiliation{Research Center of Materials Nanoarchitectonics (MANA),  
National Institute for Materials Science (NIMS), Tsukuba 305-0047, Japan
}

\date{November 24, 2025}

\begin{abstract}
We review recent theoretical progress on the charge dynamics of doped carriers in high-temperature cuprate superconductors. Advances in this field have clarified that doped charges in cuprates exhibit remarkably  rich collective behavior, governed by the combined effects of strong electronic correlations, the intrinsic layered crystal structure, and long-range Coulomb interaction. First, the emergence of acousticlike plasmons has been firmly established through quantitative analyses of resonant inelastic x-ray scattering (RIXS) spectra based on the $t$-$J$-$V$ model---an extension of the conventional $t$-$J$ model that incorporates the layered crystal structure and the long-range Coulomb interaction  $V$. These acousticlike plasmons arise near the in-plane momentum $\qp=(0,0)$ and possess characteristic energies far below the well-known $\sim$ 1 eV optical plasmon.  This behavior is found to be universal across both hole- and electron-doped cuprates, including multilayer  systems. Second,  in electron-doped cuprates,  a pronounced tendency toward $d$-wave bond-charge order develops near $\qp=(0.5\pi, 0)$, as revealed by resonant x-ray scattering and RIXS. As a result, the charge dynamics acquires a dual structure, in which low-energy bond-charge excitations coexist with relatively high-energy plasmons. Third, analogous signatures of charge-order tendency have also been reported in hole-doped cuprates. However, a direct application of the $d$-wave bond-charge-order  framework fails to account for experimental observations. Similarly, the charge-stripe order in La-based  cuprates remains unresolved within existing theoretical approaches. 
\end{abstract}

\maketitle

\section{Introduction}
In high-temperature cuprate superconductors, the superconducting phase emerges in close proximity to the antiferromagnetic phase, and the central role of spin dynamics has long been  recognized.  Spin excitations were intensively investigated already in 1990s \cite{thurston89,rossat-mignod91}, and the concept of spin-fluctuation-mediated superconductivity has since been the major theoretical scenario  for the high-$T_{c}$ mechanism \cite{scalapino12}. On the other hand, it is mobile charge carriers that actually form Cooper pairs. Therefore, a full understanding of the mechanism of high-temperature superconductivity necessarily requires not only elucidating spin dynamics but also establishing a comprehensive picture of charge dynamics \cite{yamase23}.  

High-$T_{c}$ cuprates are doped Mott insulators. In hole-doped cuprates, the antiferromagnetic phase persists up to about 5 \%, while in electron-doped compounds it extends 10--15 \%. The superconducting phase appears beyond this region, reaching a maximal $T_{c}$ around 16 \%  doping and disappearing near 20--25 \%. Since this optimal doping is not far from the antiferromagnetic phase, one might naturally expect that spin dynamics would dominate the low-energy physics. However, once carriers are doped into the system, the nearest-neighbor spin interaction inevitably induces bond-charge dynamics, as originally envisaged in resonating-valence-bond (RVB) theory \cite{anderson87,suzumura88,kotliar88,fukuyama98}.  Recent theoretical study \cite{zafur24} revealed that the spectral weight of bond-charge excitations extends over a much wider energy range and can even exceed that of spin excitations, except in the vicinity of the antiferromagnetic instability, where spin fluctuations concentrate around $\qp=(\pi, \pi)$ and exhibit nearly divergent behavior. Hence at least around the optimal doping ($\sim 16$ \%), bond-charge dynamics should be considered equally important as spin dynamics in describing the electronic property of cuprates. 

In addition to the bond-charge excitations, there exist the usual on-site charge excitations, which we sharply distinguish from bond-type to emphasize their different physical origins. It is well known that such local charge degrees of freedom have a strong tendency toward phase separation \cite{emery90,hellberg97,hellberg99} when the long-range Coulomb interaction is neglected. Although the $t$-$J$ and Hubbard models provide minimal frameworks for describing doped Mott insulators \cite{anderson87}, incorporating the long-range Coulomb interaction term $V$ represents a natural and realistic extension. Indeed, the resulting $t$-$J$-$V$ model successfully captures the essential features of charge dynamics observed in cuprates, as we will discuss below. 

Historically, in contrast to the extensive understanding of spin dynamics \cite{birgeneau06,fujita12}, detailed information on charge excitations in the plane of in-plane momentum $\qp$ and energy transfer $\omega$ became available much later, with the advent of advanced experimental technologies such as resonant x-ray scattering (RXS) and resonant inelastic x-ray scattering (RIXS) \cite{ament11,degroot24}. Consequently, the comprehensive characterization of charge dynamics is a relatively recent achievement, enabling quantitative comparisons between experiment and theory. Notably, theoretical calculations based on the $t$-$J$-$V$ model have proven remarkably successful in reproducing many key aspects of the observed spectra. 

In this review, we classify the charge dynamics in cuprates into four representative categories---i)  charge dynamics around the in-plane momentum $\qp=(0,0)$ in both electron- and hole-doped cuprates; ii) charge-order tendency near $\qp=(0.5\pi, 0)$ in electron-doped cuprates; iii) charge-order tendency around  $\qp=(0.6\pi, 0)$ in hole-doped cuprates; and iv) spin-charge stripe order in La-based cuprates. In our current understanding, the first two phenomena are  largely understood theoretically, whereas the latter two remain controversial. By thoroughly analyzing the first two, we hope to obtain a hint that may help reveal the mechanisms underlying the more elusive charge order in hole-doped and stripe-ordered cuprates. 

\section{Principal outcome from \boldmath{$t$}-\boldmath{$J$}-\boldmath{$V$} model}

\subsection{Model and theoretical framework}
We begin by defining the model and outlining the large-$N$ formalism used to analyze the charge dynamics exclusively. Our study is based on the $t$-$J$ model on a square lattice, extended to include interlayer hopping and the long-range Coulomb interaction. The Hamiltonian reads 
\be
H=-\sum_{i, j, \sigma}t_{ij}\tilde{c}_{i \sigma}^{\dagger} \tilde{c}_{j \sigma} + J \sum_{\bra i, j \ket} \left(\bold{S}_{i}\cdot \bold{S}_{j} - \frac{1}{4} n_{i} n_{j} \right) + \frac{1}{2}\sum_{i \neq j}V_{i j} n_{i} n_{j} \,,
\label{tJV}
\ee
where $\tilde{c}_{i \sigma}^{\dagger}$ and $\tilde{c}_{j \sigma}$ are creation and annihilation operators of electron with spin $\sigma (=\uparrow, \downarrow)$ at site $i$, respectively, and are defined in the restricted Hilbert space that forbids double occupancy of electrons on any  lattice site. $\bold{S}_{i}$ is the spin operator and $n_{i}=\sum_{\sigma} \tilde{c}_{i \sigma}^{\dagger} \tilde{c}_{i \sigma}$ is the density operator. The hopping amplitude $t_{i j}$ includes nearest- and next-nearest-neighbor in-plane hopping $t$ and $t'$, respectively, and interlayer hopping $t_{z}$. The superexchange $J$ acts only between nearest-neighbor sites within a plane, and the interlayer exchange is neglected as it is much smaller than $J$ \cite{thio88}. The long-range Coulomb interaction  $V_{ij}$ extends over a  three-dimensional lattice. 

Although cuprates are quasi-two-dimensional systems, it is essential to employ a layered three-dimensional model because the long-range Coulomb interaction produces a significant momentum dependence of the plasmon energy along the $q_{z}$ direction   \cite{grecu73,fetter74,grecu75}. 

The treatment of the $t$-$J$-$V$ model [\eq{tJV}] is highly nontrivial due to the local constraint that forbids double occupancy. To address this, we employ a large-$N$ technique in the path integral representation using Hubbard operators \cite{foussats02}. In this approach, the spin degrees of freedom are generalized from two to $N$ components, and physical quantities are systematically expanded in powers of $1/N$. The major advantage of this method is that it treats all possible charge excitations on an equal footing \cite{bejas12,bejas14}, allowing us to focus on the pure charge channel exclusively. 

Leaving the complete formalism to the Appendix in Ref.~\cite{yamase21a}, we here provide key steps to analyze the charge dynamics. At leading order, the electron dispersion is given by 
\be
\varepsilon_{\vk} = \varepsilon_{\vk}^{\parallel}  + \varepsilon_{\vk}^{\perp} \,,
\label{xik}
\ee
where the in-plane and out-of-plane components are given, respectively, by
\be
\varepsilon_{\vk}^{\parallel} = -2 \left( t \frac{\delta}{2}+\Delta \right) (\cos k_{x}+\cos k_{y})-
4t' \frac{\delta}{2} \cos k_{x} \cos k_{y} - \mu \,,\\
\label{Epara}
\ee
\be
\varepsilon_{\vk}^{\perp} = - 2 t_{z} \frac{\delta}{2} (\cos k_x-\cos k_y)^2 \cos k_{z}  \,. 
\label{Eperp}
\ee
The in-plane momenta $k_x$ and $k_y$ and the out-of-plane momentum $k_z$ are measured 
in units of $a^{-1}$ and $d^{-1}$, respectively, where $a$ is the in-plane lattice constant and $d$ is the interlayer spacing. Although these dispersions resemble those of noninteracting electrons, the hopping integrals $t$, $t'$, and $t_z$ are renormalized by the factor of $\delta/2$, where $\delta$ is the carrier doping concentration---this comes from a mean-field-type treatment of Hubbard operators in the $t$-term. In addition, a bond-field $\Delta$ is included in \eq{Epara}, which comes from the Hubbard-Stratonovich transformation in the $J$-term. For a given doping $\delta$, the bond-field $\Delta$ and the chemical potential $\mu$ are determined self-consistently from 
\bea
&&\Delta = \frac{J}{4N_s N_z} \sum_{\vk} (\cos k_x + \cos k_y) n_F(\varepsilon_\vk) 
{\label {Delta}} \; , \\ 
&&(1-\delta)=\frac{2}{N_s N_z} \sum_{\vk} n_F(\varepsilon_\vk)\,, 
\eea
where $n_F(\varepsilon_\vk)$ is the Fermi distribution function; $N_s$ and $N_{z}$ denote the number of sites per layer and the number of layers, respectively. 

Fluctuations around the mean fields are described by a 6-component bosonic field, $\delta X_a$ with $a=1, \dotsc, 6$.  $\delta X_1$ describes on-site charge fluctuations and $\delta X_2$ fluctuations of a Lagrange multiplier related to non-double occupancy constraint at any lattice site; $\delta X_3$ and $\delta X_4$ ($\delta X_5$ and $\delta X_6$) are real (imaginary) parts of bond-charge fluctuations along the $x$ and $y$ direction, respectively. Therefore, charge fluctuations included in the $t$-$J$-$V$ model are described by a 6 $\times$ 6 bosonic propagator. At order of $1/N$ we obtain   
\be
[D_{ab}(\vq,\mathrm{i}\omega_n)]^{-1} 
= [D^{(0)}_{ab}(\vq,\mathrm{i}\omega_n)]^{-1} - \Pi_{ab}(\vq,\mathrm{i}\omega_n)\,,
\label{dyson}
\ee
where $a, b=1, \dotsc, 6$; $\vq$ is a three-dimensional wavevector and $\nu_n$ is a bosonic Matsubara frequency. The bare propagator is given by 
\begin{eqnarray}
[D^{(0)}_{ab}({\bf q},\mathrm{i}\omega_{n})]^{-1}= N \left(
 \begin{array}{cccccc}
\frac{\delta^2}{2} \left[ V(\vq)-J(\vq)\right]
& \frac{\delta}{2} & 0 & 0 & 0 & 0 \\
   \frac{\delta}{2}  & 0 & 0 & 0 & 0 & 0 \\
   0 & 0 & \frac{4}{J}\Delta^{2} & 0 & 0 & 0 \\
   0 & 0 & 0 & \frac{4}{J}\Delta^{2} & 0 & 0 \\
   0 & 0 & 0 & 0 & \frac{4}{J}\Delta^{2} & 0 \\
   0 & 0 & 0 & 0 & 0 & \frac{4}{J}\Delta^{2} 
 \end{array}
\right).
\label{D0}
\end{eqnarray}
where  $J(\vq) = \frac{J}{2} (\cos q_x +  \cos q_y)$. The long-range Coulomb interaction in momentum space, $V(\vq)$, appropriate for a layered system, is expressed as \cite{becca96}
\be
V(\vq)=\frac{V_c}{A(q_x,q_y) - \cos q_z} \,,
\label{LRC}
\ee
where $V_c= e^2 d(2 \epsilon_{\perp} a^2)^{-1}$ and 
\be
A(q_x,q_y)=\alpha (2 - \cos q_x - \cos q_y)+1 \,,
\label{Aq}
\ee
with $\alpha=\frac{\tilde{\epsilon}}{(a/d)^2}$ and $\tilde{\epsilon}=\epsilon_\parallel/\epsilon_\perp$. Here, $\epsilon_\parallel$ and $\epsilon_\perp$ denote the 
dielectric constants parallel and perpendicular to the planes, respectively, and $e$ is the electric charge of electrons. The form of $V(\vq)$ follows from solving Poisson's equation on the lattice \cite{becca96}. 

The bosonic self-energy is 
\begin{eqnarray}
&& \Pi_{ab}(\vq,\mathrm{i}\omega_n)
            = -\frac{N}{N_s N_z}\sum_{\vk} h_a(\vk,\vq,\varepsilon_\vk-\varepsilon_{\vk-\vq}) 
            \frac{n_F(\varepsilon_{\vk-\vq})-n_F(\varepsilon_\vk)}
                                  {\mathrm{i}\omega_n-\varepsilon_\vk+\varepsilon_{\vk-\vq}} 
            h_b(\vk,\vq,\varepsilon_\vk-\varepsilon_{\vk-\vq}) \nonumber \\
&& \hspace{25mm} - \delta_{a\,1} \delta_{b\,1} \frac{N}{N_s N_z}
                                       \sum_\vk \frac{\varepsilon_\vk-\varepsilon_{\vk-\vq}}{2}n_F(\varepsilon_\vk) \; , 
\label{Pi}
\end{eqnarray}
where the $h_a(\vk,\vq,\nu)$ are $6$-component vertices and are given by  
\bea
 &&h_a(\vk,\vq,\nu) = \left\{
                   \frac{2\varepsilon_{\vk-\vq}+\nu+2\mu}{2}+
                   2\Delta \left[ \cos\left(k_x-\frac{q_x}{2}\right)\cos\left(\frac{q_x}{2}\right) +
                                  \cos\left(k_y-\frac{q_y}{2}\right)\cos\left(\frac{q_y}{2}\right) \right];
                                                   \right. \nonumber \\
               && \hspace{10mm} \left. 1; -2\Delta \cos\left(k_x-\frac{q_x}{2}\right); -2\Delta \cos\left(k_y-\frac{q_y}{2}\right);
                         2\Delta \sin\left(k_x-\frac{q_x}{2}\right);  2\Delta \sin\left(k_y-\frac{q_y}{2}\right)
                 \right\} \, .
\label{vertex-h}
\eea
Here the dependence on $k_z$ and $q_z$ enters only through $\epsilon_{\vk-\vq}$ in the first column and the other columns contain the in-plane momentum only. 

The charge excitation spectrum is obtained by analytical continuation 
\be
\mathrm{i}\omega_n \rightarrow \omega + \mathrm{i} \Gamma\,,
\label{gamma-ch}
\ee
with $\Gamma \rightarrow +0$, and by evaluating the imaginary part of $D_{ab}(\vq, \omega)$. 

When $J=0$, the bond field vanishes ($\Delta=0$), and only the usual on-site charge fluctuations remain active, reducing  $D_{ab}$ to a $2\times2$ matrix ($a,b=1,2$). $D_{11}$ describes usual charge-charge correlations; 
$D_{22}$ and $D_{12}$ correspond, respectively, to fluctuations associated with the non-double-occupancy condition and correlations between non-double-occupancy condition and charge-density fluctuations. 

For a finite $J$, the bond-charge fluctuations---the remaining $4 \times 4$ matrix---activate. Consequently full charge correlations are described by six components, that is, both on-site charge and bond-charge fluctuations coexist in a realistic situation.

Although each lattice site in the $t$-$J$-$V$ model represents a Cu atom in the CuO$_2$ plane, the effects of O atoms are implicitly included because the model is derived from the three-band Hubbard model in the strong coupling limit \cite{fczhang88}. 
Thus, while the $2\times 2$ sector of $D_{ab}$ may be associated with  Cu-site charge fluctuations and the $4 \times 4$ bond sector ($a,b=3$--$6$) with O-site charge fluctuations, this distinction is not strict owing to strong Cu-O hybridization forming the Zhang-Rice singlet \cite{fczhang88}. 

In this section, we present results for the parameters $J/t=0.3$ and $t'/t=0.30$ 
which are appropriate for electron-doped cuprates \cite{yamase15b}. 
The number of layers we take are 30, which is sufficient for convergence. 
We take $t_z = 0.1 t$, for which we checked that the Fermi surface topology  
remains the same as for $t_z=0$ in the relevant doping range. 
For the long-range Coulomb interaction [\eq{LRC}], we adopt 
$d/a=1.5$ (Ref.~\cite{misc-d}) with $a=4$~{\AA}. The dielectric constants are chosen 
as $\epsilon_\parallel=4 \epsilon_0$ and $\epsilon_\perp=2 \epsilon_0$ with $\epsilon_0$ being the dielectric constant in vacuum, consistent with experimental estimates \cite{timusk89}, although the precise value is not universal across theoretical studies \cite{becca96,prelovsek99}---those values yield $V_{c}=34$ eV and $\alpha=4.5$ in Eqs.~(\ref{LRC}) and (\ref{Aq}). 
We focus mainly on $\delta=0.15$, corresponding to the nearly optimal doping in electron-doped cuprates, which facilitates a direct comparison with RIXS experiments \cite{ishii14,wslee14}. All energy-related quantities are expressed in units of $t$.  A realistic estimate of $t/2$  (Ref.~\cite{misc-t}) in cuprates is approximately 350--500 meV (Ref.~\cite{hybertsen90}),

\subsection{Plasmon excitations}
\begin{figure}[t]
\centering
\includegraphics[width=14cm]{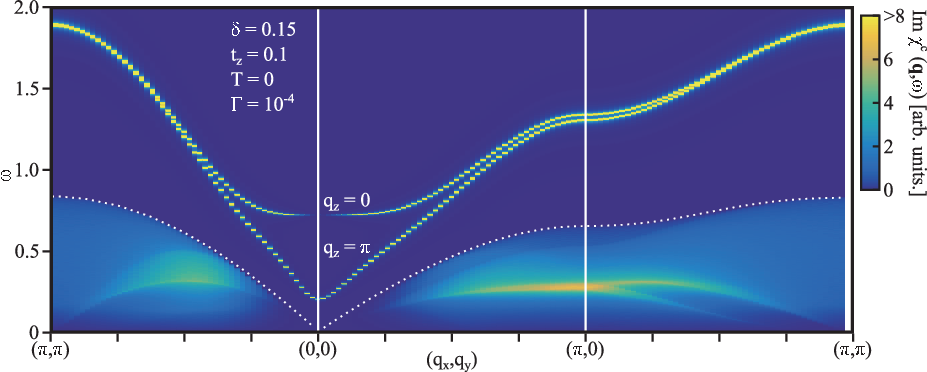}
\caption{Spectral weight of the density-density correlation function Im$\chi^{c}(\vq,\omega)$ in the plane of excitation energy $\omega$ and in-plane momentum $\qp$ along the high-symmetry path $(\pi,\pi)$--$(0,0)$--$(\pi,0)$--$(\pi,\pi)$ for $q_z=0$ and $\pi$. 
The dotted line marks the upper boundary of the particle-hole continuum for $q_{z}=0$. 
The spectral intensity of the plasmon mode is much stronger than that of the continuum and is truncated at a value of 8 to enhance the contrast of the background continuum. Adapted from Ref.~\cite{greco16} (\copyright\, 2016 American Physical Society). 
}
\label{map}
\end{figure}

As shown in Ref.~\cite{foussats02}, the usual charge-charge correlation function is expressed as 
\be
{\rm Im} \chi^{c}(\vq, \omega) = N \left( \frac{\delta}{2} \right)^{2} {\rm Im} D_{11}(\vq, \omega) \,.
\ee
Figure~\ref{map} presents a spectral intensity map of the imaginary part of the charge-charge correlation function, Im$\chi^{c}(\vq,\omega)$, in the plane of excitation energy $\omega$ and in-plane momentum $\qp$ along the symmetry direction $(\pi,\pi)$--$(0,0)$--$(\pi,0)$--$(\pi,\pi)$. Below $\omega \sim 0.8$, one observes a broad particle-hole continuum originating from individual charge excitations. This continuum shows only weak dependence on $q_z$, except that the spectral weight is slightly enhanced along $(\pi,\pi)$--$(0,0)$--$(\pi,0)$ for $q_z=\pi$ compared with $q_z=0$. In \fig{map}, we display the continuum for $q_z=0$. The absence of pronounced spectral weight near zero energy indicates that there is  no charge-order tendency associated with on-site charge degrees of freedom.  At higher energies, a sharp and intense feature emerges for  $q_z=0$. This mode corresponds to a particle-hole bound state lying above the continuum, namely the plasmon excitation. The plasmon energy at $\qp=(0,0)$ is approximately $\omega_{p}=0.7$. Its dispersion well described by 
\be
\omega(\qp, q_z=0)=\omega_{p} + a_2 q_{\parallel}^{2} + \cdots \, .
\label{plasmon}
\ee
where the ellipsis denotes higher-order terms in $\qp$. The dispersion appears almost flat near $\qp=(0,0)$, implying that the coefficient $a_2$ is quite small. The coefficient of $a_2$ can be reasonably approximated by the expression derived for a homogeneous electron gas  \cite{mahan}, $a_{2}=(3/10) v_{F}^{2}/\omega_{p}$, where $v_{F}$ is the Fermi velocity. In the present context, one may interpret $v_{F}$ as an average in-plane Fermi velocity. Because the bare hopping integrals $t$, $t'$, and $t_z$  in Eqs.~(\ref{Epara}) and (\ref{Eperp}) are renormalized by the factor of $\delta/2$, the effective velocity becomes significantly reduced at $\delta=0.15$. This renormalization provides the primary reason why the plasmon dispersion appears so flat near $\qp=(0,0)$ in \fig{map}. 

The plasmon dispersion changes dramatically when $q_{z}$ becomes finite.  As a representative example, we show the dispersion for $q_{z}=\pi$ in \fig{map}. While the plasmon dispersion for $q_{z}=\pi$ remains nearly identical to that for $q_z=0$ at large in-plane momenta, it softens substantially near $\qp=(0,0)$ and exhibits a pronounced V-shaped dispersion there. This strong $q_{z}$ dependence highlights the inherently three-dimensional character of the plasmon mode in the layered cuprates, despite their overall quasi-two-dimensional electronic structure.

\subsection{Bond-charge excitations}
We identify three major types of bond-charge excitations: $d$-wave bond-charge ($d$bond),  $s$-wave bond-charge ($s$bond), and $d$-wave charge-density-wave ($d$CDW)---also known as flux phase. To define corresponding bond-charge susceptibility, two possible schemes can be considered: i) the projection of $D_{ab}$ onto the eigenvectors associated with the respective bond-charge operators, 
and ii)  the projection of $D_{ab}^{-1}$ onto those same eigenvectors. 
Although the former definition may appear natural, it generally includes collective contributions from on-site charge fluctuations originating in the $2 \times 2$ subspace of $D_{ab}$. Their contributions obscure the intrinsic character of the bond-charge fluctuations. By contrast, the latter definition---based on $D_{ab}^{-1}$---eliminates such contamination from the on-site charge sector. Therefore we adopt the following definition for the bond-charge susceptibilities: 
\bea
&&\chi_{d{\rm bond}}^{-1} (\vq,\omega)=\frac{1}{N}\left(D_{33}^{-1}+D_{44}^{-1}-2 D_{34}^{-1}\right)\,, \label{chi-dbond}\\
&&\chi_{s{\rm bond}}^{-1} (\vq,\omega)=\frac{1}{N} \left(D_{33}^{-1}+D_{44}^{-1}+2 D_{34}^{-1} \right)\,, \\
&&\chi_{d{\rm CDW}}^{-1} (\vq,\omega)=\frac{1}{N}\left(D_{55}^{-1}+D_{66}^{-1}-2 D_{56}^{-1} \right)\,. 
\eea
We note that $\chi_{d{\rm bond}}$ corresponds to $\chi_{d{\rm PI}}$ in Ref.~\cite{bejas12} 
and $\chi_d$ in Ref.~\cite{yamase15b}.

\begin{figure}[t]
\centering
\includegraphics[width=9cm]{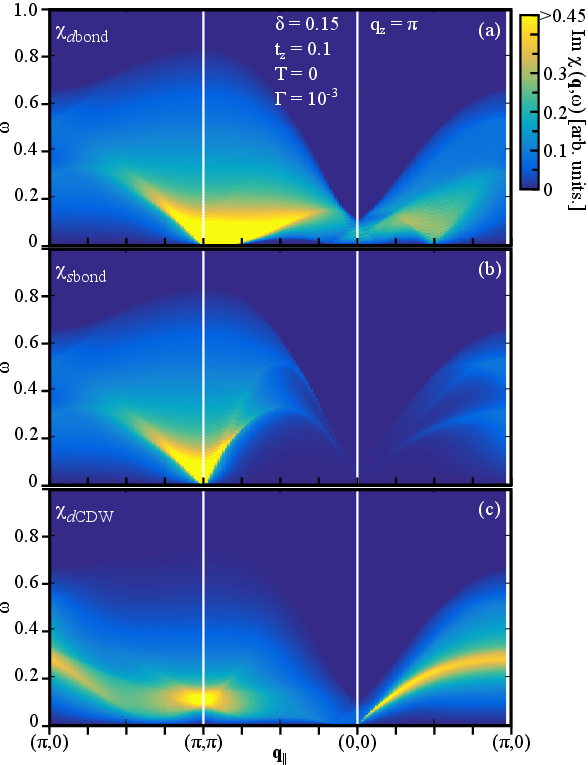}
\caption{Spectral weight maps of the bond-charge susceptibilities, 
(a) $\chi_{d{\rm bond}}$, 
(b) $\chi_{s{\rm bond}}$, and (c) $\chi_{d{\rm CDW}}$ in the plane of in-plane momentum $\qp$ and excitation energy $\omega$ along the symmetry axes $(\pi,0)$--$(\pi,\pi)$--$(0,0)$--$(\pi,0)$.
The out-of-plane momentum is $q_z=\pi$. Adapted from Ref.~\cite{bejas17}  (\copyright\, 2017 American Physical Society). 
}
\label{qw-map-D-1-pi}
\end{figure}

The excitation spectra of $\chi_{d{\rm bond}}(\vq,\omega)$, $\chi_{s{\rm bond}}(\vq,\omega)$, and  $\chi_{d{\rm CDW}}(\vq,\omega)$ are shown in \fig{qw-map-D-1-pi} along the symmetry axes for $q_z=\pi$. Before examining each mode individually, we first discuss the overall characteristics visible in \fig{qw-map-D-1-pi}. 

All three susceptibilities exhibit positive spectral weight, even though off-diagonal components of $D_{ab}(\vq,\omega)$ contain negative contributions \cite{bejas17}. At the representative doping level $\delta=0.15$, which lies close to several bond-charge instabilities, substantial spectral weight appears in the low-energy region $\omega \lesssim 0.2$. At higher energies, the spectra become broad and diffusive, indicative of incoherent excitations. These overall features show little dependence on $q_z$; similar results are obtained even in the purely two-dimensional case (see Fig.~7 in Ref.~\cite{bejas17}). 

The $\chi_{d{\rm bond}}$ spectrum displays strong low-energy intensity near $\vq_{\parallel}=0.8(\pi,\pi)$. This spectral weight is associated with the leading soft mode and it grows as the system approaches the bond-charge instability at the critical doping $\delta_c = 0.129$. Along the path from $(0,0)$ to $(\pi,0)$,  the spectral weight remains sizable, and the excitation energy decreases toward  $\vq_\parallel=(0.5\pi,0)$. This subleading mode may correspond to the charge-order tendency observed in RXS experiments \cite{da-silva-neto15}, as was first pointed out in  Ref.~\cite{yamase15b}.

The $\chi_{s{\rm bond}}$ susceptibility exhibits a low-energy dispersion centered around $\vq_\parallel=(\pi,\pi)$, reflecting the system's proximity to the corresponding instability at $\delta_c = 0.111$. The spectral weight disperses upwards with a characteristic V-shaped structure and gradually loses intensity as $\omega$ increases. In contrast to the $d$-wave case, no significant charge-order tendency is found along the $(0,0)$--$(\pi,0)$ direction. Instead, two faint dispersive features are visible along this direction, reaching $\omega \approx 0.2$ and $0.4$ near $\qp=(0.5\pi,0)$, respectively. These   weak structures reflect subtle particle-hole excitation processes rather than coherent collective modes. 
 
The $\chi_{d{\rm CDW}}$ spectrum exhibits strong spectral weight at $\vq_\parallel =(\pi,\pi)$ around $\omega = 0.1$. This energy decreases continuously with reduced doping and vanishes at the critical value $\delta_c=0.093$, where the $d$CDW instability occurs. Interestingly, a distinct gapless dispersion branch appears along the $(0,0)$--$(\pi,0)$ direction, extending up to $\omega \approx 0.3$ at $\vq_\parallel=(\pi,0)$. This mode is not a collective excitation associated  with the $d$CDW; rather, it arises from a local minimum in the real part of the denominators of $\chi_{d{\rm CDW}}$, producing a peak structure of the single-particle origin. Along the $(0,0)$--$(\pi,\pi)$ path, this dispersive feature is less pronounced, indicating an asymmetric character of $\chi_{d{\rm CDW}}$. A similar dispersing feature is observed along the $(\pi,0)$--$(\pi,\pi)$ direction, where it merges into the intense spectral weight located at $\vq_\parallel=(\pi,\pi)$ and $\omega \approx  0.1$.

\section{Experimental tests}
\subsection{Charge dynamics around \boldmath{$\qp=(0,0)$}}
\begin{figure}[b]
\centering
\includegraphics[width=16cm]{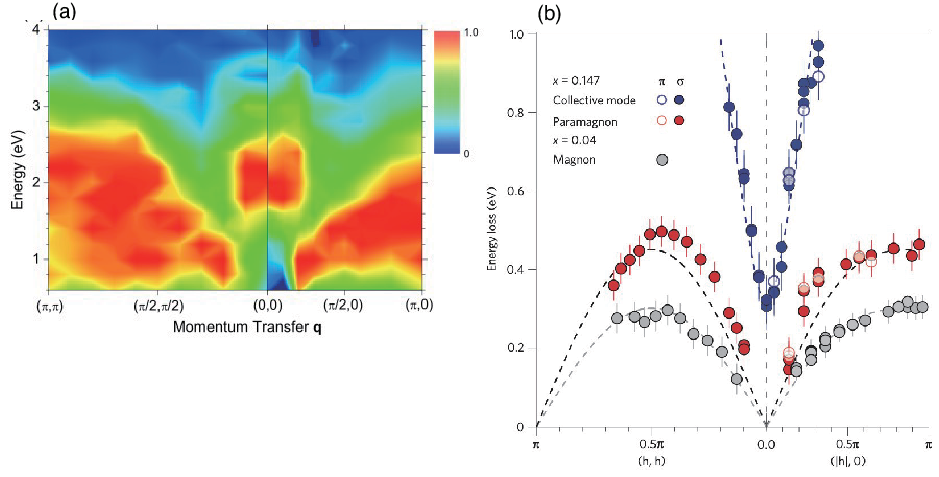}
\caption{(a) Map of charge excitations in NCCO with x=0.15. Although the spectral weight is broad, a V-shaped distribution may be  recognized around the zone center. The feature near $\omega=2$ eV at $\qp=(0,0)$ corresponds  to the charge transfer Mott gap and is irrelevant to the present work. Adapted from Ref.~\cite{ishii05}  (\copyright\, 2005 American Physical Society). (b) The V-shaped dispersion of charge excitation (blue points) near $\qp=(0,0)$ in NCCO with x=0.147. Note the vertical scale in (b) differ from that in (a). Red and gray points denote  magnetic excitations and are outside the scope of this review. Adapted from Ref.~\cite{wslee14}  (\copyright\, 2014 Springer Nature).   
}
\label{ishii-V}
\end{figure}

A characteristic V-shaped dispersion of charge excitations was first observed in electron-doped cuprates ${\rm Nd_{2-x}Ce_{x}CuO_{4}}$ (NCCO) with x=0.15,  as a broad feature in RIXS spectra as shown in \fig{ishii-V}(a) \cite{ishii05}. These excitations were initially interpreted as incoherent particle-hole excitations dressed by strong electron correlations. Similar charge excitation spectra were reported independently by two groups in 2014 [\fig{ishii-V}(b)] \cite{ishii14,wslee14}. One group, the same that first reported the feature, reaffirmed their earlier interpretation of incoherent excitations and further argued that these modes could not be collective in nature, such as plasmons \cite{ishii17}. The other group, however,  proposed a contrasting scenario: since the same signal was not observed in hole-doped cuprates, they suggested that the observed excitations represent a collective mode associated with a hidden quantum critical point specific to electron-doped cuprates \cite{wslee14}. This idea was later reinforced by subsequent measurements \cite{dellea17}.

A third and unifying interpretation was later proposed from theoretical studies of the layered $t$-$J$-$V$ model calculations as we already review in \fig{map} \cite{greco16}, where they extend the standard $t$-$J$ model by incorporating both the layered structure of cuprates and the long-range Coulomb interaction $V$. Within this framework, the V-shaped dispersion observed near the zone center can be naturally identified as an acousticlike plasmon that emerges at finite $q_{z}$ with an energy gap at $\qp=(0,0)$ proportional to the interlayer hopping $t_{z}$. In contrast, the $q_{z}=0$ limit corresponds to the well-known optical plasmon observed long ago by electron energy-loss spectroscopy (EELS) \cite{nuecker89,romberg90}. This interpretation thus offers a unified view linking the early EELS results (optical plasmons) and the more recent RIXS observations (acousticlike plasmons), both understood as manifestations of the same underlying collective mode but a different $q_{z}$ value (see \fig{map}). 

The $t$-$J$-$V$ model yielded two key predictions \cite{greco16,greco19}. i) If the V-shaped feature arises from the incoherent particle-hole excitations \cite{ishii05,ishii14,ishii17}, it should display almost no $q_{z}$ dependence. ii) If it is of plasmonic origin, the mode energy should rapidly decrease with increasing $q_{z}$ from zero for small $\qp$. 

In 2018, Hepting {\it et al.} \cite{hepting18} confirmed the latter prediction experimentally, reporting that in NCCO with x=0.15 and 0.175 the plasmon energy indeed decreases with increasing $q_{z}$ from  $q_{z}=0$. Interestingly, the observed dispersion resembled an acoustic mode in that the plasmon energy appeared to approach zero linearly as  $\qp \rightarrow {\bf 0}$, suggesting that the predicted small gap was below the experimental resolution. To clarify this points, Hepting {\it et al.} \cite{hepting22} later performed RIXS measurements on the electron-doped infinite-layer compound ${\rm Sr_{0.9}La_{0.1}CuO_{2}}$ (SLCO), where  the spacing between adjacent CuO$_{2}$ planes is smaller than the in-plane lattice constant $a$. Consequently, the interlayer hopping $t_{z}$ is expected to be relatively large, and thus the plasmon gap should be experimentally accessible. In fact, the RIXS spectra revealed a clear gap of about 120 meV \cite{hepting22}, in excellent agreement with the theoretical prediction that the plasmon gap scales with $t_{z}$, thereby providing strong support for the plasmon interpretation \cite{greco16}. 

\begin{figure}[ht]
\centering
\includegraphics[width=8cm]{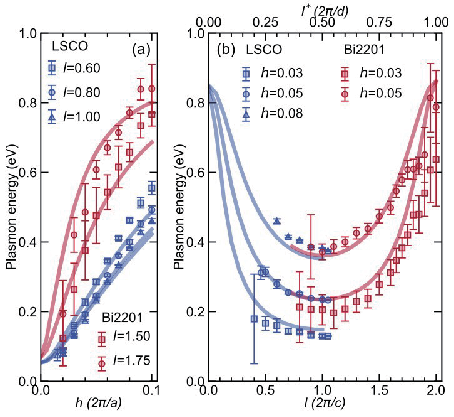}
\caption{(a) In-plane dispersion of plasmons along the $q_{x}$ ($h$) for several $q_{z}$ ($l$) values.  (b) Out-of-plane dispersion along the $q_{z}$ ($l$) for several choices of $q_{x}$ ($h$) values. Blue points correspond to LSCO with x=0.16 and red ones to Bi2201. Solid curves are $t$-$J$-$V$ model calculations using  $t/J=0.3$, $t/2=350$ meV, $t'=-0.20 t$ for LSCO and $t'=-0.35t$ for Bi2201; $V_{c}=18.9$ eV (52.5 eV), $\alpha=3.47$ (8.14), and $\Gamma=0.20$ (0.29) for LSCO (Bi2201). The larger values of $V_{c}$ and $\alpha$ in Bi2201 arise mainly from its large interlayer spacing $d$. The upper limit of the interlayer hopping $t_{z}$ is $0.01t$ in LSCO. Adapted from Ref.~\cite{nag20}  [\copyright\, 2020 The Author(s)].    
}
\label{fit-pl}
\end{figure}

A further theoretical implication of the plasmon scenario is its universality: the acousticlike plasmon  mode should exist not only in electron-doped cuprates but also in hole-doped ones. This has been confirmed experimentally by Nag {\it et al.} \cite{nag20} for ${\rm La_{2-x}Sr_{x}CuO_{4}}$ (LSCO) with x=0.16 and ${\rm Bi_{2}Sr_{1.6}La_{0.4}CuO_{6+\delta}}$ (Bi2201), and Singh {\it et al.} \cite{singh22a} for LSCO with x=0.12. Figure~\ref{fit-pl} is quantitative comparisons between RIXS and the $t$-$J$-$V$ model calculations, revealing good agreement for both in-plane ($q_{x}$) and out-of plane ($q_{z}$) dispersion by choosing parameter sets appropriate to each compound.  In addition, the qualitative trend in spectral intensity is also reproduced \cite{nag20,hepting22}.

Theoretically the doping dependence of the plasmon gap was clarified in Ref.~\cite{greco16}. A comparison with experiments has been done in Ref.~\cite{hepting23}. It showed a nice agreement up to doping rate 16 \% and a deviation from the theoretical prediction above that. As possible reasons, Ref.~\cite{hepting23} discussed three possibilities: i) a problem of sample quality in the overdoped region in LSCO, ii) overestimation of correlation effects of the $t$-$J$-$V$ model, and iii) nonplanar Cu and O orbitals effect might be responsible for this discrepancy.

So far, most RIXS experiments have focused on single-layer or infinite-layer cuprates, i.e., systems with one CuO$_{2}$ plane per unit cell. Recently, attention has turned to multi-layer cuprates, motivated by the empirical trend that the superconducting transition temperature $T_{c}$ increases with the number of CuO$_{2}$ planes up to three \cite{iyo07}. Exploiting charge dynamics in these systems is therefore of particular interest. 

For double-layer cuprates, two plasmon branches,  $\omega_{\pm}$, were predicted long ago within the electron gas model \cite{fetter74,griffin89}. This prediction is valid even in electron liquid system like cuprates \cite{yamase25}. Recent random-phase-approximation (RPA)  \cite{yamase25,sellati25} and bilayer $t$-$J$-$V$ model \cite{yamase26} calculations have shown that the experimentally observed plasmon mode \cite{bejas24} corresponds well to the lower $\omega_{-}$ branch. 

In three-layer cuprates, an additional $\omega_{3}$ mode is theoretically predicted alongside $\omega_{\pm}$ \cite{griffin89}. Although calculations beyond the electron-gas model are not yet available, recent experiments \cite{nakata25} suggest that the measured spectra can be mainly attributed to the $\omega_{-}$ mode, with possible contributions also from $\omega_{3}$ and $\omega_{+}$ modes. Understanding how these multiple charge excitation branches evolve with the number of layers---and how they might relate to the enhancement of $T_{c}$---remains an intriguing open question for future studies.

\subsection{Charge dynamics around \boldmath{$\vq_{\parallel}=(0.5\pi,0)$} in electron-doped cuprates}
RXS probes the equal-time charge correlation function, which is defined as for $d$-wave bond-charge fluctuations 
\be
S(\vq)=\frac{1}{\pi} \int_{-\omega_c}^{\omega_c}  {\rm d}\omega \, {\rm Im} \chi_{d{\rm bond}}(\vq,\omega) 
\left[ n_B(\omega)+1 \right], 
\label{Sq}
\ee
where $n_B(\omega)=1/({\rm e}^{\omega/T}-1)$ is the Bose distribution function and $T$ is temperature. The cutoff energy $\omega_c$ is introduced for a later convenience and, in the standard case of equal-time correlation, we have $\omega_c=\infty$ theoretically. The $d$-wave bond-charge susceptibility, obtained within the leading order of the large-$N$ expansion [see \eq{chi-dbond}], reads 
\be
\chi_{d{\rm bond}}(\vq, \omega)= \frac{(8\Delta^2/J)^{-1}} {1-2J\Pi_{d}(\vq, \omega)} \,, 
\label{chid}
\ee
which becomes exact in the limit of large $N$. Here $\Delta$ denotes the mean-field value of the bond field introduced in \eq{Epara}. Among various charge components, only the $d$-wave bond charge exhibits a clear softening along $(0,0)$--$(\pi, 0)$ direction, as shown in \fig{qw-map-D-1-pi}. We therefore first focus on this component. In the next subsection, we will discuss the tendency of charge order around $\qp=(\pi, \pi)$. 

Since the three dimensionality is not important in contrast to Sec.~III.~A, we do not consider $q_{z}$ dependence. Hence $\vq=\qp$ in what follows. In this case, the present formalism yields results consistent with those obtained by dynamical density-matrix renormalization-group method \cite{tohyama15,greco17} and exact diagonalization method \cite{merino03,bejas06}.

The $d$-wave polarization function $\Pi_{d}(\vq,\omega)$ in \eq{chid} is given by 
\be
\Pi_{d}(\vq, \omega) = - \frac{1}{N_{s}}\;
\sum_{\vk}\; \gamma^2(\vk) \frac{f(\epsilon_{\vk + \vq/2}) 
- f(\epsilon_{\vk - \vq/2})} 
{\epsilon_{\vk + \vq/2} - \epsilon_{\vk- \vq/2}-\omega - {\rm i}\Gamma}\,,
\label{Pid}
\ee
where the $d$-wave form factor $\gamma(\vk)=(\cos k_x - \cos k_y)/2$ characterizes the $d$-wave symmetry of the bond-charge order, and $\Gamma$ is an infinitesimally small damping factor. In the limit of $\vq={\bf 0}$, $\chi_{d{\rm bond}}(\vq,\omega)$ would be reduced to the 
electronic nematic susceptibility \cite{yamase04b} associated with a $d$-wave Pomeranchuk instability \cite{yamase00a,yamase00b,metzner00}. 

\begin{figure} [th]
\centering
\includegraphics[width=5cm]{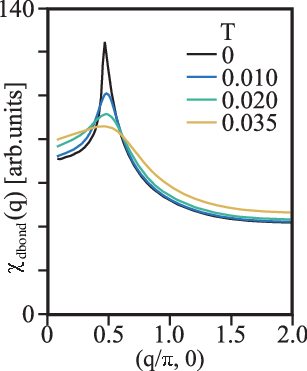}
\caption{
$\vq$ dependence of static $d$-wave bond-charge susceptibility $\chi_{d{\rm bond}}(\vq)$ for several temperatures at a fixed doping level of $\delta=0.13$. The enhancement of $\chi_{d{\rm bond}}(\vq)$ with decreasing temperature signals the development of $d$-wave bond-charge correlations.  
Adapted from Ref.~\cite{yamase19}  (\copyright\, 2019 American Physical Society).}
\label{Gamma0001}
\end{figure}

For the present analysis, we take $J=0.3$ and $t'=0.3$ in the Hamiltonian (\ref{tJV}), representing typical parameters for electron-doped cuprates \cite{bejas14}. The $V$ term is replaced by the nearest-neighbor Coulomb interaction, and the precise value of $V$ is unimportant as long as it prevents phase separation. We choose $\Gamma=10^{-4}$ in \eq{Pid}, a sufficiently small value. Figure \ref{Gamma0001} displays the static $d$-wave bond-charge susceptibility $\chi_{d{\rm bond}}(\vq)=\chi_{d{\rm bond}}(\vq,\omega=0)$ as a function of $\vq$ for several  temperatures at $\delta=0.13$. Because of the $d$-wave form factor [see \eq{Pid}], $\chi_{d{\rm bond}}(\vq)$ exhibits a $4\pi$ periodicity along the $q_{x}$ direction, and thus the momentum range is restricted to  $0 \leq q_x \leq 2\pi$ in  \fig{Gamma0001}. As the temperature decreases,  pronounced peaks appear at $\vq=(\pm 2 \pi Q_{\rm co}, 0)$ and $(0,\pm 2 \pi Q_{\rm co})$ with $Q_{\rm co}\approx 0.25$, indicating a tendency toward a charge-order tendency. However, the static susceptibility does not diverge,  and  thus the charge order remains short ranged. 

\begin{figure} [th]
\centering
\includegraphics[width=8cm]{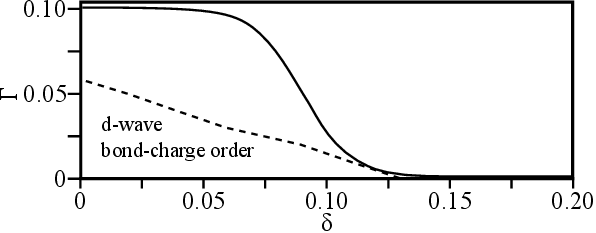}
\caption{
Doping dependence of the damping parameter $\Gamma$  (solid line) and 
the zero-temperature phase boundary of the $d$-wave bond-charge order (dashed line).  Adapted from Ref.~\cite{yamase19}  (\copyright\, 2019 American Physical Society). }
\label{Gamma-delta}
\end{figure}

Phenomenologically, the parameter $\Gamma$ represents the broadening of charge excitation spectrum. In light of this, we may reasonably assume $\Gamma$ depends on $T$, $\delta$, $\vq$, and $\omega$. Because our primary interest lies in the temperature and doping dependence of $S(\vq)$, 
we may introduce the $T$- and $\delta$-dependences in $\Gamma$. 
At finite temperature, a leading correction linear in $T$ is expected \cite{misc-MFL},     
\be
\Gamma(\delta,T) = \Gamma(\delta) + \alpha T \,.
\label{Gamma-T}
\ee
The doping dependence can be inferred from neutron scattering data \cite{motoyama07}, which show that the antiferromagnetic correlation length increases markedly below $\delta \approx 0.10$. Concomitantly, quasiparticles are expected to becomes more heavily damped in this regime. On the basis of this phenomenology, we adopt a simple parameterization,  
\be
\Gamma(\delta) =0.001 + 0.05\left[ 1- \tanh \left(\frac{\delta-0.09}{0.02}\right) \right] \,,
\label{Gamma-form}
\ee
where $\Gamma$ rises steeply below $\delta \approx 0.10$ (\fig{Gamma-delta}). 

The dashed line in \fig{Gamma-delta} indicates the $T=0$ phase boundary of the $d$-wave bond-charge order. For infinitesimal and doping-independent $\Gamma$, the instability occurs at $\delta_{c} \approx 0.125$ at $T=0$. Increasing $\Gamma$ suppresses the ordered phase, so that only short-range charge fluctuations survive above the dashed lines in \fig{Gamma-delta}. 
As a result, we have only charge fluctuations associated with the $d$-wave 
bond-charge order for doping above the dashed line in \fig{Gamma-delta}. Note that as expected from \fig{qw-map-D-1-pi}(a), the $d$-wave bond-charge instability may occur around $\qp\approx 0.8(\pi,\pi)$, not $Q_{\rm co}=0.25$, for $\Gamma=+0$; this feature is analyzed in the next subsection. While the choice of the absolute value of $\Gamma$ is rather arbitrary in \eq{Gamma-form}, 
we choose it to have no charge instabilities even at low-doping rate at $T=0$ (solid line in 
\fig{Gamma-delta}), so that our calculations are performed in the paramagnetic phase in the entire doping region. 
The temperature coefficient $\alpha=9$ in \eq{Gamma-T} is chosen after verifying that 
the qualitative results remain unchanged for $\alpha=3$ and $6$. While the precise functional form of $\Gamma(\delta, T)$ is not essential, it must increase with temperature and with decreasing doping to suppress charge instabilities in the low-doping region.

\begin{figure}[htb]
\centering
\includegraphics[width=8cm]{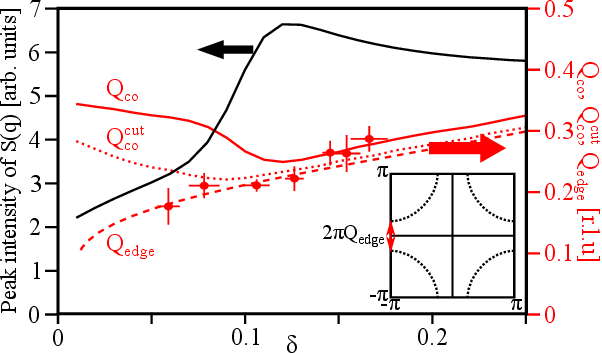}
\caption{Peak intensity of $S(\vq)$ [see \eq{Sq}] with $\omega_c=\infty$ 
and the corresponding momenta $Q_{\rm co}$, $Q_{\rm co}^{\rm cut}$, and  $Q_{\rm edge}$ 
as a function of hole doping $\delta$ at $T=0$. $Q_{\rm co}$ denotes the peak position of  
$S(\vq)$ obtained with $\omega_c=\infty$, while $Q_{\rm co}^{\rm cut}$ is that determined with a cutoff $\omega_c=0.05$. $Q_{\rm edge}$ is defined in the inset. 
Solid circles represent the experimental data from Ref.~\cite{da-silva-neto16}.  
Adapted from Ref.~\cite{yamase19}  (\copyright\, 2019 American Physical Society).}
\label{delta-Sq}
\end{figure}

As shown in \fig{delta-Sq}, the spectral intensity $S(\vq)$ is strongly suppressed with decreasing doping below $\delta \approx 0.10$, consistent with experimental observations  \cite{da-silva-neto16}. This suppression originates from the rapid increase in $\Gamma$ (\fig{Gamma-delta}). If $\Gamma$ were constant, the peak intensity of $S(\vq)$ would continue to grow upon lowering  $\delta$. 

\begin{figure}[htb]
\centering
\includegraphics[width=12cm]{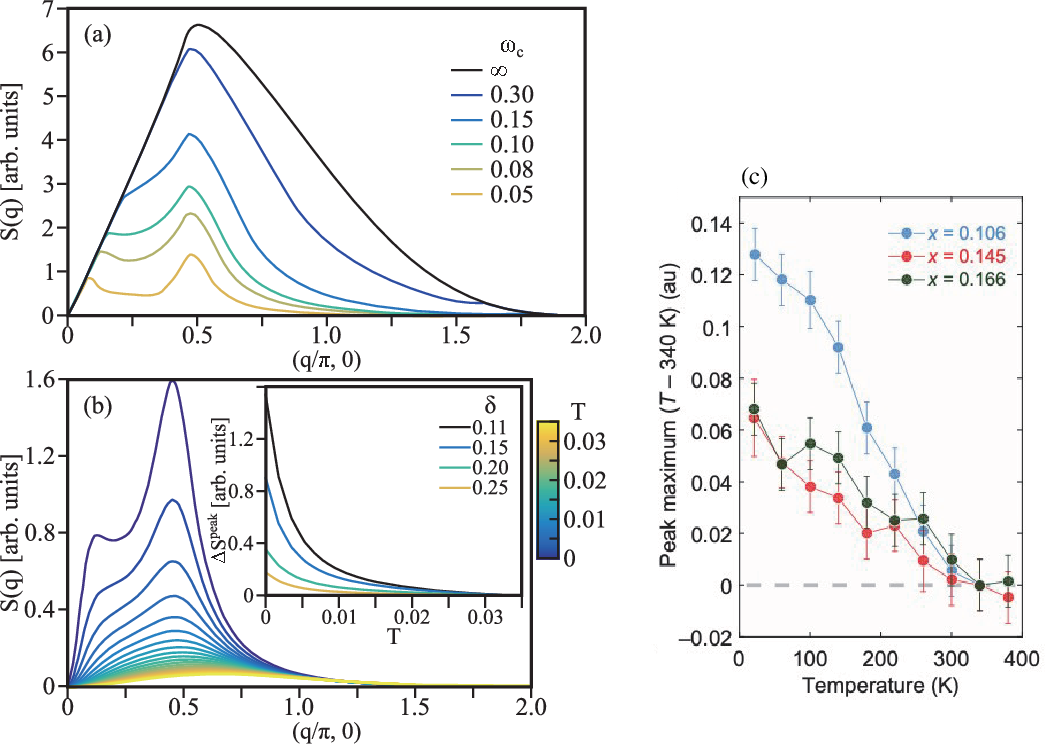}
\caption{(a) $\vq$ dependence of $S(\vq)$ for several choices of the cutoff energy at $\delta=0.13$ and $T=0$.
(b) $\vq$ dependence of $S(\vq)$ with a cutoff energy $\omega_c=0.05$ for 
various temperatures at $\delta=0.13$. 
The inset shows the temperature dependence of the peak intensity of $\Delta S$ 
for several doping levels, obtained after subtraction of intensity at $T=0.035$. 
Adapted from Ref.~\cite{yamase19}  (\copyright\, 2019 American Physical Society). 
(c) Temperature evolution of the peak intensity from $T=340$ K at several doping levels. Adapted from Ref.~\cite{da-silva-neto16}  [\copyright\, 2016 The Author(s)]. }
\label{Sq_cut}
\end{figure}

$S(\vq)$ exhibits a peak at $\vq=(\pm 2\pi Q_{\rm co}, 0)$ and $(0, \pm 2\pi Q_{\rm co})$ (see Figs.~\ref{Gamma0001} and \ref{Sq_cut}). The peak positions are plotted in \fig{delta-Sq} as a function of doping, together with $Q_{\rm edge}$---the distance between the Fermi surface edges across $\vk=(\pi,0)$ (see the inset of \fig{delta-Sq}). The peak structure arises from particle-hole scattering processes characterized by $Q_{\rm edge}$. Hence $Q_{\rm co}$ corresponds to $Q_{\rm edge}$ at least down to $\delta\approx 0.10$, 
although it becomes slightly larger because $S(\vq)$ is energy-integrated [\eq{Sq}].  
Below $\delta \approx 0.10$, $Q_{\rm co}$ deviates significantly from $Q_{\rm edge}$, reflecting the strong damping that blurs Fermi-surface features.

The peak of $S(\vq)$ in \fig{Sq_cut}(a) is broad even at $T=0$. 
This broadness is not due to the finite $\Gamma$ of Eqs.~(\ref{Gamma-T}) and 
(\ref{Gamma-form}), since we can check that $S(\vq)$ remains broad even for $\Gamma=10^{-4}$ [see Fig.~1(b) in Ref.~\cite{yamase19}]. Instead, it  originates from the energy integration in \eq{Sq}. Indeed, when 
the cutoff energy $\omega_{c}$ is reduced, the peak sharpens substantially [\fig{Sq_cut}(a)]. 
The sharp peak for low $\omega_{c}$ reflects the short-range $d$-wave bond-charge order, which is otherwise smeared by higher-energy contributions. Hence it is meaningful to analyze $S(\vq)$ with a low cutoff, as in recent RIXS studies \cite{da-silva-neto18}. 

Figure~\ref{Sq_cut}(b) shows $S(\vq)$ for $\omega_c = 0.05$ at various temperatures. 
A broad structure at high $T$ evolves into a distinct peak below $T \sim 0.02$, signaling the growth of short-range $d$-wave bond-charge correlations. To track its evolution, we define 
$\Delta S(\vq) = S(\vq; T) - S(\vq; T=0.035)$, where $T=0.035$ ($\approx$ 300--400 K) serves as a background. The temperature dependence of the peak intensity $\Delta S^{\rm peak}$, shown in the inset of \fig{Sq_cut}(b), increases at lower $T$ and lower doping, essentially consistent with experiments  [\fig{Sq_cut}(c)].  The peak position  $Q_{\rm co}^{\rm cut}$ obtained from low $\omega_{c}$ data follows $Q_{\rm edge}$ down to low doping and begins to rise only below $\delta \approx 0.1$ (\fig{delta-Sq}). If we do not consider seriously the experimental data at $\delta=0.059$, where the existence of a peak at $Q_{\rm co}$ in Fig. 2(B) of Ref.~\cite{da-silva-neto16} is unclear, the experimentally observed decrease of $Q_{\rm co}$ with lowing doping \cite{da-silva-neto16}---nearly saturating below $\delta \approx 0.10$---is well reproduced by our theory (\fig{delta-Sq}). In particular, the low-energy component $Q_{\rm co}^{\rm cut}$ shows quantitative agreement over a wide doping range.

In the low-energy window [\fig{Sq_cut}(b)], the peak narrows markedly below $T \sim 0.01 (\sim 100$ K), which becomes comparable to the experimental data. This suggests that our theoretical spectra overemphasize high-energy contributions in \eq{Sq}. Two factors likely contribute: (i) the experimental spectral weight above $\sim 0.1$ eV is weaker than in theory, leading to a narrower experimental line shape; (ii) RIXS measurements \cite{da-silva-neto18} indicate that the observed signal contains not only change but also magnetic excitations, extending from 200--700 meV at $Q_{\rm co}$ and intensifying below 300 K---effects not included in the present model. 

Our results also capture other essential features observed in RIXS experiments. In Ref.~\cite{da-silva-neto18}, the charge-order signal originates from energies below 60 meV, while our calculated sharp feature appears for $\omega_c \approx 0.05 - 0.1$, which may correspond to 40--100 meV. 

The experimental finding that charge and magnetic excitations share a similar energy scale  \cite{da-silva-neto18} is also naturally explained. As shown in Sec.~II, $d$-wave bond-charge order comes from the spin exchange interaction, i.e., the $J$-term in the $t$-$J$ model. If the charge dynamics originated instead from local on-site charge excitations, its characteristic energy would be much higher, as in plasmon excitations \cite{greco16,bejas17,greco19}. Thus both bond charge and magnetic dynamics emerge on the same scale of $J$, reflecting their common microscopic origin. 

Finally, the experimental $Q_{\rm co}$ shows little temperature dependence \cite{da-silva-neto16}. In our calculations, when focusing on the low-energy regime [\fig{Sq_cut}(b)], $Q_{\rm co}^{\rm cut}$ is similarly temperature-independent below $T \sim 0.01 ( \sim 100$ K).

\subsection{Possible charge ordering wth \boldmath{$\qp=(\pi, \pi)$}} 
As shown in \fig{qw-map-D-1-pi}, the strong intensity is concentrated around $(\pi, \pi)$ in a low-energy, implying a  
strong tendency of bond-charge instability around $\qp=(\pi,\pi)$. However, RIXS cannot measure such a large momentum region. Experimental tests of the predicted signal around $(\pi,\pi)$ are left to future studies. 

As a caveat, we note here that a signal around $(\pi,\pi)$ behaves in a very special way for $d$-wave bond-charge fluctuations. As shown in Fig.~\ref{qw-map-D-1-pi}, the $d$-wave bond-charge order develops at  $\vq_{1} = (\pm q_1, \pm q_1)/\sqrt{2}$, while the instability at  $\vq_{2} = (\pm 2 \pi Q_{\rm co}, 0)$ and $(0, \pm 2 \pi Q_{\rm co})$ represents the second-leading one. Despite this hierarchy, the peak structure of $S(\vq)$, as well as that of $\chi_{d{\rm bond}}(\vq)$, exhibits markedly different behavior at these two wave vectors. Upon lowering temperature, the peaks at $\vq_{2}$ sharpen significantly (\fig{Gamma0001}), whereas the spectral feature around $\vq_{1}$ remains typically broad. A discernible peak at $\vq_{1}$ emerges only in the immediate vicinity of the onset temperature of the charge instability. This distinctive contrast was analyzed in detail in Ref.~\cite{yamase15b}.

\begin{figure} [th]
\centering
\includegraphics[width=7cm]{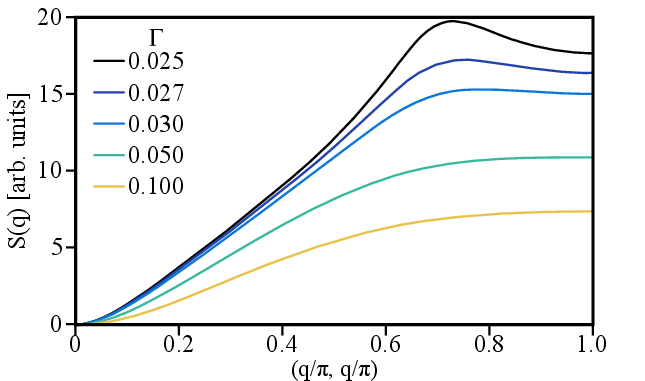}
\caption{$S(\vq)$ along the $(0,0)$--$(\pi,\pi)$ direction for various choices of $\Gamma$ at $T=0$ and $\delta=0.08$; the cutoff energy is $\omega_c = \infty$. Adapted from Ref.~\cite{yamase19}  (\copyright\, 2019 American Physical Society). 
}
\label{Sq-diagonal}
\end{figure}

In the present work, we consider a relatively large value of $\Gamma$. Under this condition, the peak structure around $\vq_{1}$ is generally suppressed and becomes visible only when the system is tuned very close to the phase boundary of the $d$-wave bond-charge order. To illustrate this point explicitly, we compute $S(\vq)$ along the $(0,0)$--$(\pi,\pi)$ direction for several values of $\Gamma$ at fixed $\delta = 0.08$ and $T = 0$, as shown in Fig.~\ref{Sq-diagonal}. For large $\Gamma$, $S(\vq)$ displays only a broad structure. A small but discernible peak around $(0.75\pi, 0.75\pi)$ develops only when $\Gamma$ is reduced to $\Gamma = 0.025$, which is very close to the phase boundary.

Because the peak structure around $\vq_{1}$ is generally absent in the presence of a large $\Gamma$, we have focused on the peak structure at $\vq_{2}$ in Sec.~III.~B. This wave vector is directly relevant to RXS measurements and to the experimental observations reported in Refs.~\cite{da-silva-neto15, da-silva-neto16, da-silva-neto18}.

\section{Perspectives on charge orders in cuprates}

\subsection{Charge dynamics around \boldmath{$\qp=(0.6\pi,0)$} in hole-doped cuprates}
Given the success in describing electron-doped cuprates with $d$-wave bond-charge order, it is natural to ask whether a similar framework can be applicable to hole-doped cuprates. However, no softening is observed along the $(0,0)$--$(\pi,0)$ direction within the present theoretical framework. In fact, the origin of the charge-order tendency in hole-doped cuprates remains controversial. 

Various theoretical frameworks have been proposed to explain the charge order observed in hole-doped cuprates. A commonly encountered difficulty is that the ordering wavevector predicted by theory is substantially smaller than that observed experimentally \cite{bejas12,allais14,meier14,yamakawa15,zeyher18}. This discrepancy appears to be mitigated when the pseudogap is explicitly incorporated \cite{atkinson15}; however, a satisfactory and widely accepted microscopic description of the pseudogap remains controversial.

Ref.~\cite{bejas12} performed a comprehensive analysis of charge instabilities in the $t$-$J$ model within a large-$N$ formalism, identifying bond-charge orders with various symmetries, including $d$-wave bond-charge order. Ref.~\cite{allais14} also proposed $d$-wave bond-charge order arising from essentially the same mechanism as discussed in Sec.~III.~B of the present work. In Ref.~\cite{meier14}, a bidirectional charge-density-wave (CDW), not bond-charge order, was discussed as being induced near the Brillouin zone edge by superconducting fluctuations inside the pseudogap phase in proximity to an antiferromagnetic quantum critical point. The bidirectional CDW, however, is not compatible with experiments \cite{kawasaki24}. 

Ref.~\cite{yamakawa15}, based on a three-orbital Hubbard model, obtained a CDW wave vector determined by neighboring hot spots, again smaller than experiments. There, both nearest-neighbor and on-site Coulomb interactions enhance the charge susceptibility, while an axial CDW is selected through triangular diagrams in the Aslamazov-Larkin vertex corrections (AL-VC). Because the AL-VC scales with spin fluctuations, strong spin fluctuations are essential, which is not supported at least for electron-doped cuprates as we have reviewed in Sec.~III. B. The importance of AL-VC was also emphasized in Ref.~\cite{zeyher18}, which studied bond-charge orders rather than a conventional CDW; axial charge fluctuations are likewise enhanced by this mechanism.

Ref.~\cite{atkinson15} employed an effective three-band model and, assuming quasistatic magnetic moments and an antiferromagnetic correlation length exceeding that of the charge order, obtained four Fermi pockets resembling those of an antiferromagnetically ordered state. These assumptions, however, are not supported experimentally. The Cu 4$s$ orbital worked to stabilize an axial CDW emerging from the pseudogap via short-range Coulomb interactions.

The doping dependence of the charge-order signal remains unresolved and was left open in Ref.~\cite{atkinson15}. In the spin-fermion analysis of Ref.~\cite{wang14}, a successive phase transitions—nematic, time-reversal symmetry breaking, and CDW—was found upon cooling, with onset temperatures increasing toward lower doping, contrary to experiments; a similar doping dependence was obtained also in Ref.~\cite{zeyher18}.

Whereas most theoretical approaches invoked spin-induced bond-charge order \cite{bejas12, allais14,zeyher18} or spin fluctuations \cite{meier14,wang14,yamakawa15}, Ref.~\cite{mishra15} cast doubt on an itinerant spin-fluctuation-mediated origin of the CDW.

The key difference from the electron-doped case is that the charge order in hole-doped cuprates is observed inside the pseudogap phase \cite{keimer15}. As recently clarified \cite{zafur24}, the detailed band dispersion around the Fermi surface strongly affects the spectral weight distribution of bond-charge excitations. In the pseudogap phase, substantial modification to the band structure is expected, creating a gaplike feature. This pseudogap effect is challenging to reproduce accurately in theoretical models. Therefore one possible reason why many calculations fail to capture the experimentally observed charge-ordering behavior is an inadequate treatment of the pseudogap physics. In other words, understanding the origin of charge order may provide crucial insight into the underlying mechanism of the pseudogap---a major open problem in high-$T_{c}$ cuprates. 

\subsection{Charge ordering in La-based cuprates}
\begin{figure}[t]
\centering
\includegraphics[width=8cm]{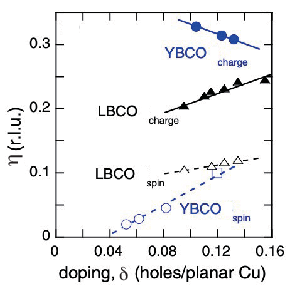}
\caption{Doping dependence of wavevector of charge and spin order in ${\rm La_{2-x}Ba_{x}CuO_{4}}$ (LBCO) and ${\rm YBa_{2}Cu_{3}O_{6+y}}$ (YBCO). The spin and charge ordering wavevectors are defined as ($\pi -2 \pi \eta_{\rm spin}, \pi)$ and $(2\pi \eta_{\rm charge}, 0)$, respectively.  Adapted from Ref.~\cite{blackburn13}  (\copyright\, 2013 American Physical Society). 
}
\label{q-delta}
\end{figure}

The doping dependence of the charge-order momentum of La-based cuprates (LBCO) is shown in \fig{q-delta}---the charge-order wavevector increases with doping. In contrast, in YBCO as a representative of other hole-doped cuprates,  the charge-order wavevector around $\qp=(0.6\pi, 0)$  decreases monotonically with increasing doping. This contrast suggests that the origin of charge order in La-cased cuprates differs from that in other hole-doped cuprates \cite{atkinson15}. Nonetheless, there are theoretical proposals that charge order in both La-based and other cuprates can be understood consistently within a stripe framework \cite{ido18,ohgoe20} or in a three-orbital model via AL-VC \cite{yamakawa15}. 

The distinct feature of La-based cuprates is that the charge order wavevector $(2\pi \eta_{\rm charge}, 0)$ is approximately twice the incommensurate magnetic wavevector $(\pi- 2 \pi \eta_{\rm spin}, \pi)$, i.e., $\eta_{\rm charge}=2\eta_{\rm spin}$ (Ref.~\cite{tranquada95}). This observation strongly indicates  that charge order in these materials is likely driven by coupling to magnetic fluctuations or spin order \cite{yamase99,huang18}.

Recently numerical studies have been revisiting this issue. For example, simulations of the Hubbard model with $t'=0$ at $\delta=1/8$ find a charge stripe ground state without $d$-wave superconductivity \cite{bxzheng17}. Results are sensitive to the value of $t'$. Ref.~\cite{jiang19} reported coexistence of stripes with superconductivity at 1/8 doping, while Ref.~\cite{ponsioen19} found charge order consistent with experiments in La-based cuprates in $-0.43 \lesssim t'/t \lesssim  -0.09$ at 1/8 doping, with superconductivity emerging for $0.18 \lesssim \delta <0.25$. Ref.~\cite{ido18} further demonstrated that the stripe period decreases with increasing doping, reproducing the experimentally observed trend, and confirmed the period-4 stripes at 1/8 doping, consistent with experiment, although the model predicted  phase separation at low doping ($0 < \delta \lesssim 0.1$) and a stripe phase extending up to $\delta \sim 0.5$. 

\subsection{Charge dynamics under spin-fluctuations}

Spin fluctuations are believed important in cuprate physics. Here we summarize the charge dynamics under spin fluctuations both experimentally and theoretically. 

According to RXS data observed in NCCO \cite{da-silva-neto16}, the authors clarified that spin fluctuations are not responsible for the charge order. But later RIXS experiments \cite{da-silva-neto18} found that there are spin fluctuations at the same wavevector as charge order, in a higher energy region. In this sense, we cannot deny any connection between spin fluctuations and the charge order. On the other hand, as clarified in La-based cuprates, spin and charge are strongly coupled with each other—the wavevector of charge $Q_{\rm charge} = 2 Q_{\rm spin}$. 

Theoretically, as reviewed in Secs.~II.~C and III.~B, the bond-charge order originates from the spin-spin (instantaneous) interaction \cite{bejas12,allais14,zeyher18,zafur24}. On the other hand, there are several proposals that spin fluctuations are responsible for a usual CDW as seen in theory associated with a quantum critical point \cite{meier14}, the spin-fermion model \cite{wang14}, and a model invoking AL-VC \cite{yamakawa15}. 

\subsection{Charge order and pseudogap}

The pseudogap is also believed to be essential for high-$T_{c}$ cuprate physics. Here we summarize possible connection of charge order and the pseudogap. 

Most of theoretical studies do not consider the pseudogap when analyzing the charge order. However, Atkinson {\it et al.} \cite{atkinson15} considered the spin-density-wave state, which can mimic the Fermi surface observed in the pseudogap. Although the pseudogap state is not an antiferromagnetic ordered phase, their obtained charge order is rather close to the experimental observation. This implies how important the pseudogap is for charge ordering. They also emphasized that the charge order is not a primary source of the pseudogap. 

A similar conclusion was obtained by calculating the electron self-energy that the charge fluctuations are not related directly with the pseudogap formation \cite{yamase24}—they lead to a Fermi liquid state in the low-energy limit, but the quasiparticle weight is substantially reduced. This might suggest that the charge fluctuations contribute constructively to the pseudogap formation because the quasiparticle weight vanishes in the pseudogap phase. However, counterintuitively, a smaller quasiparticle weight requires a stronger pseudogap self-energy. This intricate interplay offers an interesting future issue. 

Experimentally, the charge order was observed inside the pseudogap phase in Y-based cuprates \cite{keimer15}. However, in Bi-based cuprates, the charge order occurs simultaneously with the pseudogap formation \cite{comin14}. This may not necessarily mean that the charge ordering is likely the origin of the pseudogap. Instead, it implies the intimate relation between the pseudogap and charge ordering across distinct materials.

\section{Conclusions}
After reviewing the principal outcome from the $t$-$J$-$V$ model in Sec.~II.~A, 
we have shown in Secs.~II.~B and III.~A that the charge dynamics around $\qp =(0,0)$ 
is quantitatively described in terms of plasmons with a finite $q_{z}$.  
In Secs.~II.~C and III.~B, we have demonstrated that the charge-order tendency observed in electron-doped cuprates 
can be captured by the $d$-wave bond-charge order, not a usual CDW. 
These agreements suggest four key implications for understanding the charge dynamics in cuprates. 
i) The $t$-$J$-$V$ model provides a minimal framework to capture the essential physics of charge dynamics in cuprates.  ii) The origin of the bond--charge order lies in the magnetic exchange interaction, i.e., $J$-term, and consequently its energy scale is determined by  $J$.  iii) Interestingly, the bond-charge order is not driven by antiferromagnetic fluctuation, but the instantaneous interaction $J$.  Antiferromagnetic fluctuations appear primarily to enhance quasiparticle damping, which in turn suppress the tendency toward charge ordering. iv) Conceptually, it is important to distinguish bond-charge order and a usual CDW. In Sec.~IV, we have reviewed the currently controversial issues. Given a belief that the underlying physics is universal across hole- and electron-doped cuprates, we hope that partial successes of the $t$-$J$-$V$ model to understand the charge dynamics will serve to develop further studies of the cuprate physics.

\acknowledgments
The author sincerely thanks numerous colleagues for their invaluable contributions and insightful discussions: M. Bejas, E. H. da Silva Neto, A. Greco, M. Hepting, B. Keimer, W. Metzner, M. Minola, A. Nag, S. Nakata, H. Suzuki, T. Tohyama, M. Zafur, R. Zeyher, Ke-Jin Zhou, and L. Zinni. In particular, the core ideas presented in this review have greatly benefited from close collaborations with M. Bejas and A. Greco. Special appreciation is extended to the Max-Planck-Institute for Solid State Research in Stuttgart for their warm hospitality. This work was supported financially by JSPS KAKENHI Grants No.~JP18K18744 and JP20H01856, and by the World Premier International Research Center Initiative (WPI), MEXT, Japan.

\vspace{5mm}
\noindent e-mail address: yamase.hiroyuki@nims.go.jp






\bibliography{main} 

\begin{thebibliography}{92}
\bibitem [{mis({\natexlab{a}})}]{misc-d}%
  \BibitemOpen
  \href@noop {} {} ({\natexlab{a}}),\ \bibinfo {note} {although CuO$_2$ planes
  shift by $(\frac{1}{2}, \frac{1}{2}, \frac{1}{2})$ in
  Nd$_{2-x}$Ce$_{x}$CuO$_{4}$, we model the actual system by neglecting such a
  shift for simplicity. Our interlayer distance $d$ is thus given by a half of
  the $c$-axis lattice constant.}\BibitemShut {Stop}%
\bibitem [{mis({\natexlab{b}})}]{misc-t}%
  \BibitemOpen
  \href@noop {} {} ({\natexlab{b}}),\ \bibinfo {note} {a factor of $1/2$ comes
  from a large-$N$ formalism where $t$ is scaled by $t/N$ and the physical
  situation corresponds to $N=2$.}\BibitemShut {Stop}%
\bibitem [{mis({\natexlab{c}})}]{misc-MFL}%
  \BibitemOpen
  \href@noop {} {} ({\natexlab{c}}),\ \bibinfo {note} {stimulated by the
  marginal Fermi liquid theory \cite{varma96}, one might wonder how our results
  are modified if $\Gamma$ also depends on $\omega$ by assuming $\Gamma \propto
  {\rm max}(T, \omega)$ at fixed density. In this case, the peak intensity and
  the peak position of $S(\vq)$ may change, but our conclusions are not
  modified.}\BibitemShut {Stop}%
\end{thebibliography}

\end{document}